\documentstyle[11pt,paspconf]{article}
\input psfig.tex

\def\lsim{\lower.5ex\hbox{$\; \buildrel < \over \sim \;$}}
\def\gsim{\lower.5ex\hbox{$\; \buildrel > \over \sim \;$}} 

\begin{document}

\title{Unified Accretion Disk Models Around Black Holes and Neutron Stars 
and Their Spectral Properties}

\author{S. K. Chakrabarti}
\affil{Tata Institute of Fundamental Research, Homi Bhabha Road, Mumbai, 400005, INDIA}

\begin{abstract}
We review the current understanding of accretion flows around compact
objects with a special emphasis on advective disks. We discuss the
influence of the centrifugal pressure supported high density region 
around compact objects (where shocks may also form) on the emitted
spectra. We show that the stationary and non-stationary spectral properties 
(such as, low and high states, transition of states, quasi-periodic oscillations, 
quiescent and rising phases of X-ray novae, etc.) of both low mass 
and supermassive black hole candidates could be 
satisfactorily explained within the framework of the analytical solution 
of the advective disks without invoking any ad hoc components such as 
Compton clouds or magnetic corona.

\end{abstract}


\section{Fundamental Properties of Advective Disks}
Our understanding of accretion processes around black holes is vastly
improved from the analysis of the global solutions which include
advection, rotation, viscosity, heating and cooling processes
(Chakrabarti, 1990; 1996a; 1996b). Central to the physics of these advective
disks is the fact that matter enters through
the horizon with a velocity equal to the velocity of light
and therefore, the accretion flow must be supersonic and hence
sub-Keplerian (Chakrabarti, 1990; 1996bc) in the nearby region. 
(This was completely ignored by standard disk models of Shakura \& Sunyaev,
1973 and Novikov \& Thorne, 1973). Thus, every flow must deviate
from a (subsonic) Keplerian disk near a black hole. The location from where the
deviation occurs as well as  the location of the sonic points depend 
on the heating and cooling effects (i.e., roughly speaking, 
on viscosity parameter and mass accretion rate). 
Further, close to a black hole, the infall time
scale is much smaller compared with the viscous time scale, and matter
plunges into the hole with roughly constant angular momentum. Thus,
the centrifugal barrier becomes stronger even for a small angular momentum
(the barrier would be infinite for a Newtonian potential) which causes the density
of matter to rise rapidly. Matter piled up behind this barrier may or 
may not produce standing shock waves depending on whether the shock conditions
are satisfied. Shock or no shock, the behavior of the enhanced 
emission of the hotter radiation from the denser region behind the centrifugal barrier
appears to be sufficiently relevant to explain the observed spectral 
properties (stationary or non-stationary) of galactic and extragalactic 
black hole candidates (Chakrabarti \& Titarchuk, 1995; Chakrabarti, 1997).

Fig. 1 shows the classification of the {\it entire} parameter space
according to the types of solutions that are prevalent around a black hole
(Chakrabarti, 1996c; see also Chakrabarti, 1989; 1990). 
In the central box, the parameter space (spanned by specific 
angular momentum $l$ and specific energy ${\cal E}$) is divided
into nine regions marked by $N$, $O$, $NSA$, $SA$, $SW$, $NSW$, $I$,
$O^*$, $I^*$. The horizontal line at ${\cal E}=1$ corresponds to the rest
mass of the flow. Surrounding this parameter space, 
\begin{figure}
\centerline{
\psfig{figure=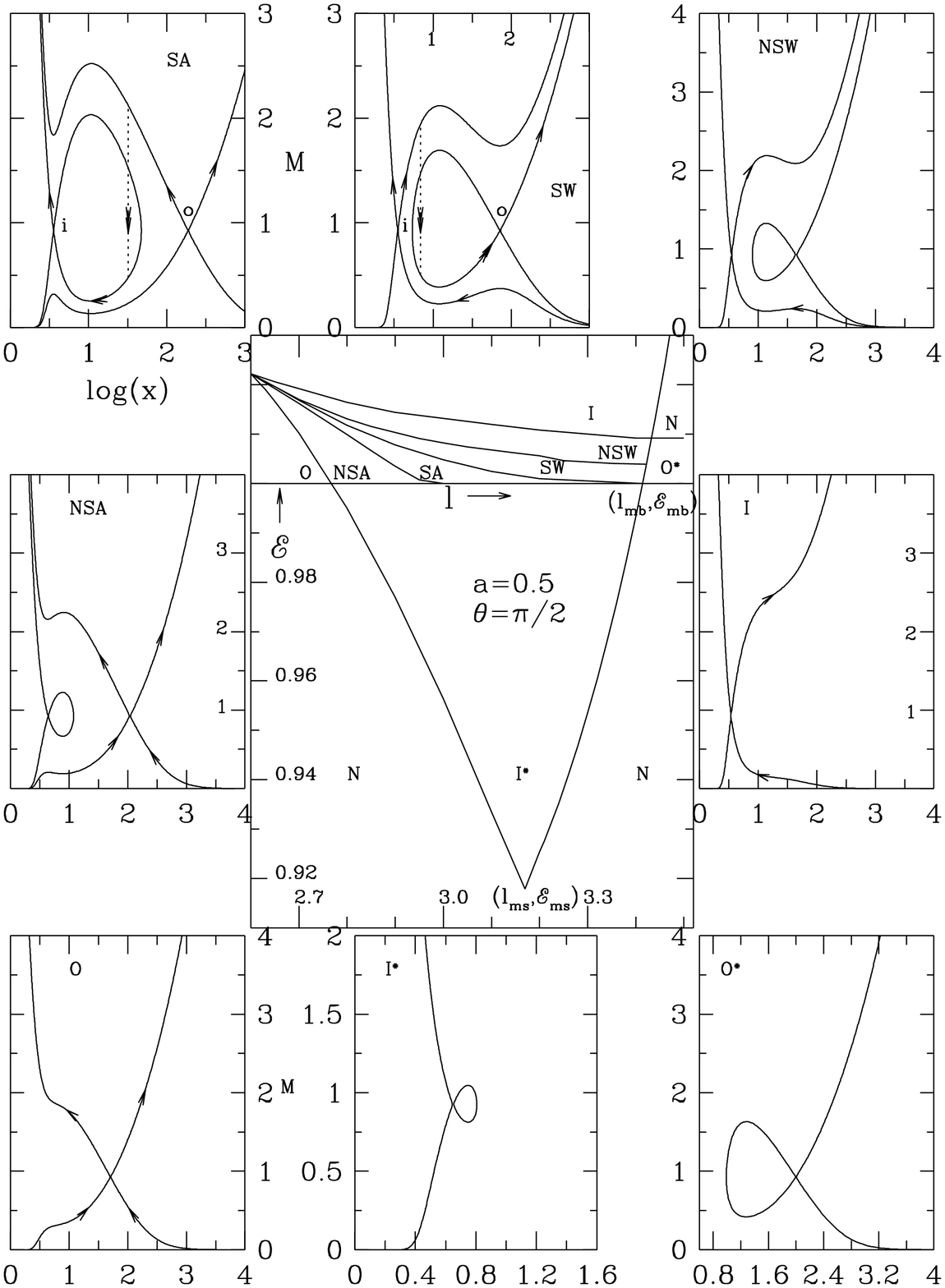,height=11.0truecm,width=11.0truecm,angle=0}}
\vspace {0.2cm}
\noindent {\small {\bf Fig. 1:} 
Classification of the {\it entire} parameter space (central box)
in the energy-angular momentum plane in terms of topological variation 
of the Kerr black hole accretion ($a=0.5$ and polytropic index $\gamma=4/3$). 
Eight surrounding boxes show the solutions from each independent region of the
parameter space. Contours are of constant entropy accretion rate ${\dot{\cal M}}$.
Vertical arrowed lines correspond to shock transitions.}
\end{figure}
{\parfillskip=0pt 
various solutions (Mach number $M=v_x/a_s$ vs. logarithmic radial distance $x$
where $v_x$ is the radial velocity and $a_s$ is the sound speed) marked
with the same notations (except $N$) are plotted. The accretion 
solutions have inward pointing arrows and the wind solutions have outward pointing arrows.
The region $N$ has no transonic solution. $E$ and $l$ are the only two 
parameters required to describe the entire inviscid global solutions. Since 
$E$ is assumed to be constant, entire energy is advected towards the hole. Thus,
these solutions are hot but inefficient radiators very similar to their
spherical counterpart (Bondi flow). In the case of neutron star
accretion, the subsonic inner boundary condition forces the flow to choose the
sub-sonic branch and therefore the energy must be dissipated at the
shock (at $x_{s1}$ or $x_{s3}$ in the notation of Chakrabarti, 1989)
outside the neutron star surface (Chakrabarti, 1989; 1990; 1996b)
unless the entire flow is subsonic.
The solutions from the region `O' has only the outer sonic point. 
The solutions from the regions $NSA$ and $SA$ have two `X' type sonic points
with the entropy density $S_o$ at the outer sonic point {\it less} than the
entropy density $S_i$ at the inner sonic point. However, flows from $SA$
pass through a standing shock  since the Rankine-Hugoniot
conditions are satisfied. The entropy generated at the shock 
$S_i-S_o$ is advected towards the black hole to enable the flow to pass
through the inner sonic point. Such solutions have been verified
by detailed numerical simulations (Chakrabarti \& Molteni, 1993;
Molteni, Lanzafame \& Chakrabarti, 1994; Molteni, Ryu \& Chakrabarti, 1996;
see also Chakrabarti et al, 1997). Rankine-Hugoniot conditions are not satisfied
for flows from the region $NSA$. Numerical simulations show
(Ryu, Chakrabarti \& Molteni, 1997; Chakrabarti et al. 1997)
that flows from this region are very unstable
and exhibit periodic changes in emission properties as they
constantly try to form stationary shocks, but fail to do so. The frequency 
and amplitude of modulation (10-50\%) of emitted X-rays have properties similar
to Quasi-Periodic Oscillations observed in black hole candidates (Dotani, 1991).
In galactic black holes, these frequencies are around 1Hz (exact number
depends on shock location, i.e., $l,\ {\cal E}$ parameters) but for 
extragalactic systems the time scale could range from a few hours
to a few days depending on the central mass ($T_{QPO} \propto M_{BH}$).
Numerous cases of QPOs are reported in the literature 
(e.g., Dotani, 1991; Halpern \& Marshall, 1996; Papadakis \& Lawrence, 1995).
In presence of cooling effects, otherwise stationary shocks from $SA$ also 
oscillate with frequency and amplitude modulations
comparable to those of QPOs {\it provided} the cooling timescale is roughly comparable to the
infall timescale in the post-shock region (Molteni, Sponholz \& Chakrabarti, 1996). 
Kilohertz oscillations on neutron stars (van der Klis, this volume) are also possible 
when the shock at $x_{s1}$ form. 
The solutions from the regions $SW$ and $NSW$ are very similar to those from
$SA$ and $NSA$. However, $S_o \geq S_i$ in these cases.
Shocks can form only in winds from the region $SW$. Shock conditions are not
satisfied in winds from the region $NSW$. This makes the $NSW$ flows
unstable as well. A flow from region $I$ has only the inner sonic
point and thus can form shocks (which require the presence of two 
saddle type sonic points) if the inflow is already supersonic 
due to some other physical processes (as in a wind-fed system).
Each solution from regions $I^*$ and $O^*$ has two sonic points (one `X' and one `O')
only and neither produces complete and global solution. The region $I^*$
has an inner sonic point but the solution does not extend subsonically
to a large distance. The region $O^*$ has an outer sonic point, but the
solution does not extend supersonically
to the horizon! When a significant viscosity is added, the closed
topology of $I^*$ opens up and then the flow joins with a cool Keplerian 
disk (C90ab; C96) which has ${\cal E} <1$. These special solutions of viscous 
transonic flows should not have shock waves. However, hot flows deviating from
a Keplerian disk or sub-Keplerian companion winds, or flows away from
an equatorial plane (Chakrabarti, 1996d) 
or, cool flows subsequently preheated by magnetic flares or irradiation
can have ${\cal E}>1$ and therefore standing shock waves. 
Note that in order to have standing shocks, one does not
require any large angular momentum. Indeed, in most of the cases, the flow
need to have $l<<l_{ms}$, the marginally stable value (Fig. 1). Although 
a flow with polytropic index $\gamma<1.5$ does not have shocks, \par}
\newpage
\begin{figure}
\centerline{
\psfig{figure=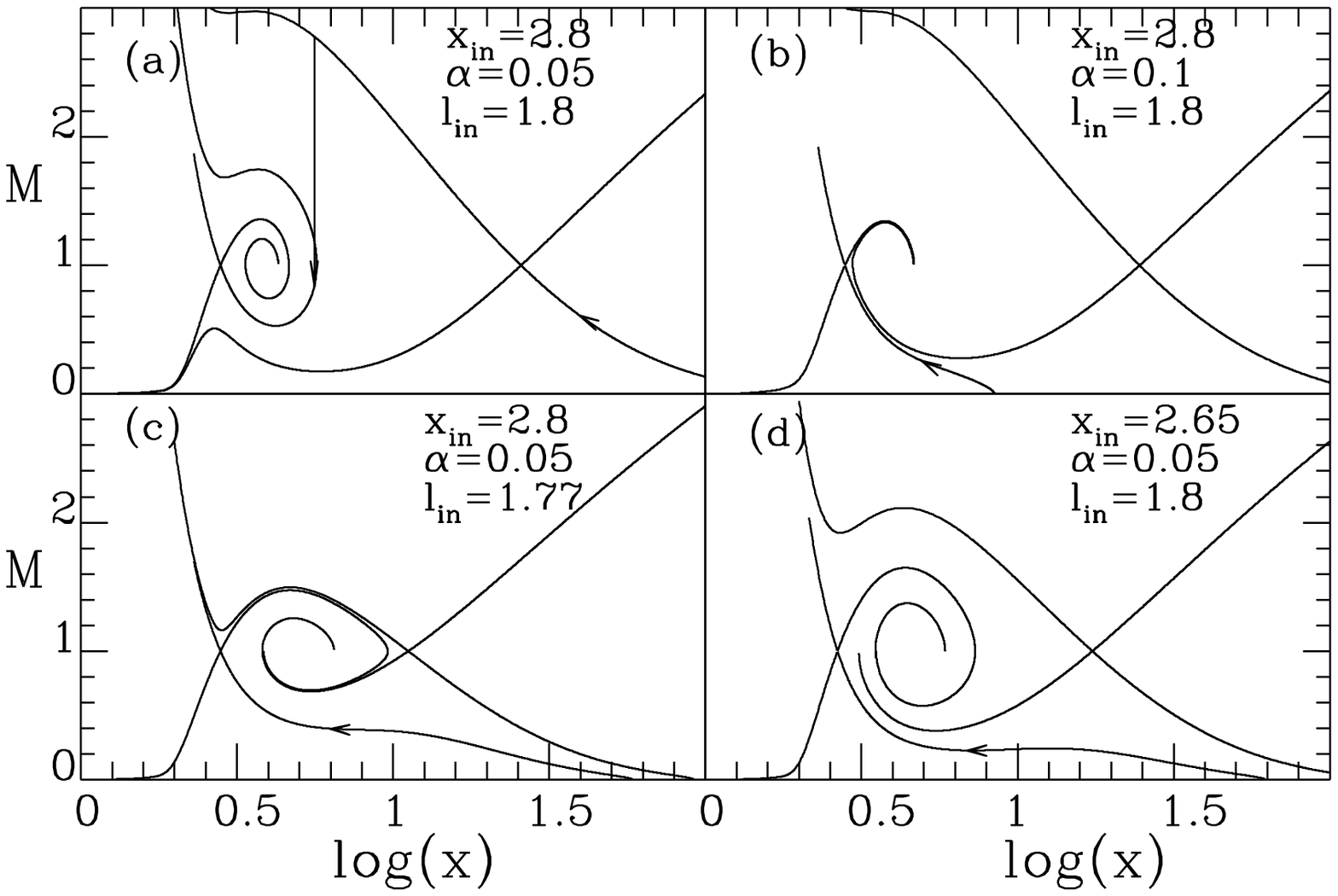,height=13.0truecm,width=13.0truecm,angle=0}}
\vspace{-7.0cm}
\noindent {\small {\bf Fig. 2:} 
Change in solution topology of a viscous flow as the three parameters are varied
(Chakrabarti, 1990). In (a), after deviating from a hot Keplerian disk the flow can pass through 
the outer sonic point, shock and the inner sonic point (shown in arrows), while
in (b-d) the stable flow can pass through the inner sonic point after it
deviates from a Keplerian disk. Density and velocity distribution in the flow close to the
black hole remain roughly the same in all these cases. Distance $x$ is measured in units of
$x_g$.}
\end{figure}
\noindent the stationary observational properties, which depend only on the 
enhanced emission from the region behind of centrifugal barrier, are not affected.

When viscosity is added, the closed topologies shown in  
Fig. 1 open up as the `O' type sonic points become spiral or nodal type.
Singularly important in this context is the non-trivial change in topology 
when each of the three free parameters are changed (Chakrabarti, 1990). 
In the context of viscous isothermal flows (These discussions are valid
for a general flow as well, see, Chakrabarti, 1996ab.)
these free parameters can be chosen to be inner sonic point $x_{in}$ 
(this replaces the specific energy parameter), the angular momentum 
on the horizon $l_{in}$ and the viscosity parameter $\alpha$.
Temperature of the disk is computed self-consistently from these parameters.
Fig. 2 shows that transition to topology in (b-d) from topology (a) 
can take place either by increase in viscosity or decrease 
$x_{in}$ or $l_{in}$. In (a), shocks are still possible, while in (b-d),
shocks do not form as the flow enters into the hole through the inner sonic point 
straight away from a Keplerian disk, but the density variation and 
emission properties remain similar to that of a shocked flow. In (b),
Keplerian disk is extended close to the horizon, while in (a)
the deviation takes place farther outside of the outer sonic point. Thus, for instance, if 
there is a vertical variation of viscosity in a Keplerian disk very far 
away from a black hole, it is possible that different layers would deviate 
from a Keplerian disk at different radial distance, and a sub-Keplerian flow
(with or without a shock) would surround a Keplerian disk as the flow approaches
the compact object.
The sub-Keplerian flow could also be contributed by companion winds in
wind-fed systems. This two component advective flow (TCAF) Model (Chakrabarti \& Titarchuk,
1995), for the first time, obviates the need of having any ad hoc Compton 
cloud or magnetic coronae required in the past to explain the power 
law components of black hole spectra. Here, one computes the properties 
of the so called `Compton cloud' self-consistently since it is a part of the inflow itself.
Extensive numerical simulations (Chakrabarti \& Molteni, 1995; Chakrabarti et al. 1997) 
of viscous advective flows verify these analytical findings.  

In Fig. 3, we show a common property of advective flows which 
`feel' centrifugal barrier close to the hole. We plot the ratio of 
radial to azimuthal velocities (solid curves) as well as the density (dashed 
curves) of the flow (in arbitrary units) as functions of the radial distance 
of three illustrative examples of solutions (Chakrabarti, 1996b). 
Here, $f=(Q_+-Q_-)/Q_+$ is chosen to be constant throughout the flow
for simplicity. For a given angular 
\begin{figure}
\vbox{
\vskip -0.5cm
\centerline{
\psfig{figure=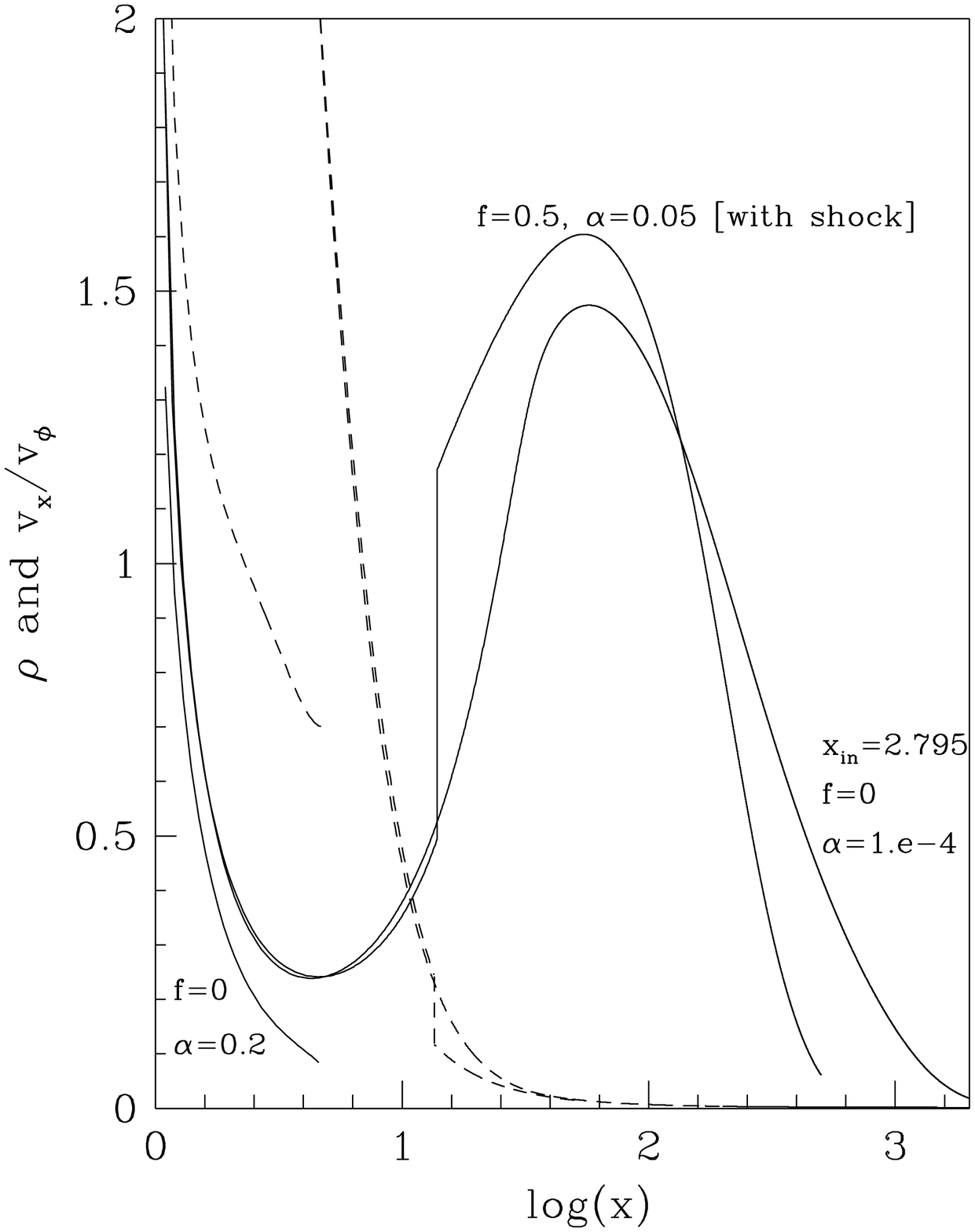,height=8.0truecm,width=8.0truecm,angle=0}}}
\vskip -0.7cm
\noindent {\small {\bf Fig. 3:} 
Ratios $v_x/v_\phi$ (solid) and densities (dashed) of three illustrative
solutions of the advective flows (Chakrabarti, 1996b). Note that the centrifugal barrier
close to the hole makes all the three solutions to
behave similarly in the region $2\lsim x \lsim 10$,
emission from which strongly determines the spectral properties of the hole.
In a strongly shocked flow the variations occur in a shorter length scale while
in a weakly shocked or shock-free flows the variations occur in an extended
region. }
\end{figure}
momentum of the flow
at the horizon, solutions without shocks (marked with $\alpha=1.e-4$ and $\alpha=0.2$)
and that with a shock ($\alpha=0.05$) have similar properties close to a black hole,
namely, all have $v_x/v_\phi <<1$ around $2 \lsim x \lsim 10$. Density and velocity
distributions
in the postshock region are similar to those in the subsonic shock-free flow. 
Thus, in some sense, all the solutions have `shocks' behind the centrifugal barrier. 
Only the length scales in which the quantities change vary. Hence, {\it qualitative} 
spectral properties of the TCAF Model does not seriously depend on whether
the shocks actually form in either or both of the Keplerian 
(which also becomes sub-Keplerian close to the hole)
and the sub-Keplerian components. However, spectral properties {\it do  depend} upon 
whether the accretion flow is of single component (such as one Keplerian
flow becoming entirely sub-Keplerian close to the hole) or two components 
(where the original Keplerian disk plus companion winds are segregated into Keplerian and
sub-Keplerian components very far away before being mixed into a single sub-Keplerian
component near the hole). This will be demonstrated below.

\section{Observational Properties of Two Component Advective Flows.}
Based upon the theoretical unstanding of the properties of Advective flows,
Chakrabarti \& Titarchuk (1995) pointed out that 
the accretion on most compact objects may be taking place in two components:
one is of higher viscosity, predominantly Keplerian (Disk Component) and is extended
till around $x_K=10x_g$ if the accretion rate is high enough to keep it thermally
and viscously stable, otherwise $x_K$ could be higher (see also, Ebisawa, Titarchuk
\& Chakrabarti, 1996). Keplerian region of the disk component supplies soft photons.
The other component (Halo Component) is predominantly sub-Keplerian (which is originated 
from the Keplerian disk far away and is contributed 
by companion winds, if present, in the case of a galactic black hole 
and by winds from numerous stars in the case of a supermassive black hole.) 
This component radiates inefficiently and therefore is hot ($\sim$ virial temperature) and
together with $x<x_K$ of the disk component they supply hot electrons which in turn
energize intercepted soft photons to produce hard component. The extent to which 
electrons cool is determined by the accretion rates in these two components. 
At least three important variations of this Model is recognized: TCAFM1-- 
In this case, the halo component forms a strong shock behind the centrifugal 
barrier: $x_s\sim 10x_g$ and puffs up and mixes up with the disk component at $x<x_s$.
TCAFM2-- The halo component does not form a shock or 
forms only a weak shock but still feels the centrifugal barrier as in TCAFM1. 
TCAFM3-- The halo component is completely devoid of angular momentum. 
The disk component deviates from a Keplerian disk at $x_K$. 
For $x<x_K$ these components mix as before. In this case, the absence of 
centrifugal barrier reduces the optical depth of the region $x<x_K$ and
it is easy to cool this region even at a low disk rate. 
A corollary of these Models is a single component model SCAFM, where
the sub-Keplerian component rate is so low that it is practically non-existent.
In SCAFM, soft photons of the Keplerian region may or may not cool the hot electrons
of its own sub-Keplerian region (for $x<x_K$) very effectively depending on $x_K$
and the spectra remains soft in most of the parameter space. Also, in this
case, the hard and soft components are always anticorrelated while observations
suggest that very often they behave independently. 
In Chakrabarti \& Titarchuk (1995), TCAFM1 is extensively studied while
other possibilities are also mentioned (see also Ebisawa et al. 1996). More detailed
study of these models are in Chakrabarti (1997). Fig. 4a schematically shows
the possible flow model based on the analytical considerations. 
Note that this generalized disk of the 1990s 
(and hopefully of the future) is really a natural combination of purely 
advecting Bondi flow of the 1950s
and purely rotating Keplerian disks of the 1970s. In Fig. 4b we show the basic
difference in the soft state spectra of neutron stars and black holes. In the soft state,
the disk rate is large and emitted soft photons completely cool the inner quasi-spherical
sub-Keplerian region. The inner boundary property on the horizon of a black hole causes the
cool (but rushing with velocity comparable to the velocity of light) electrons 
to Comptonize a fraction of these soft photons due to direct momentum transfer 
(as opposed to random 
\begin{figure}
\vbox{
\vskip -0.5cm
\centerline{
\psfig{figure=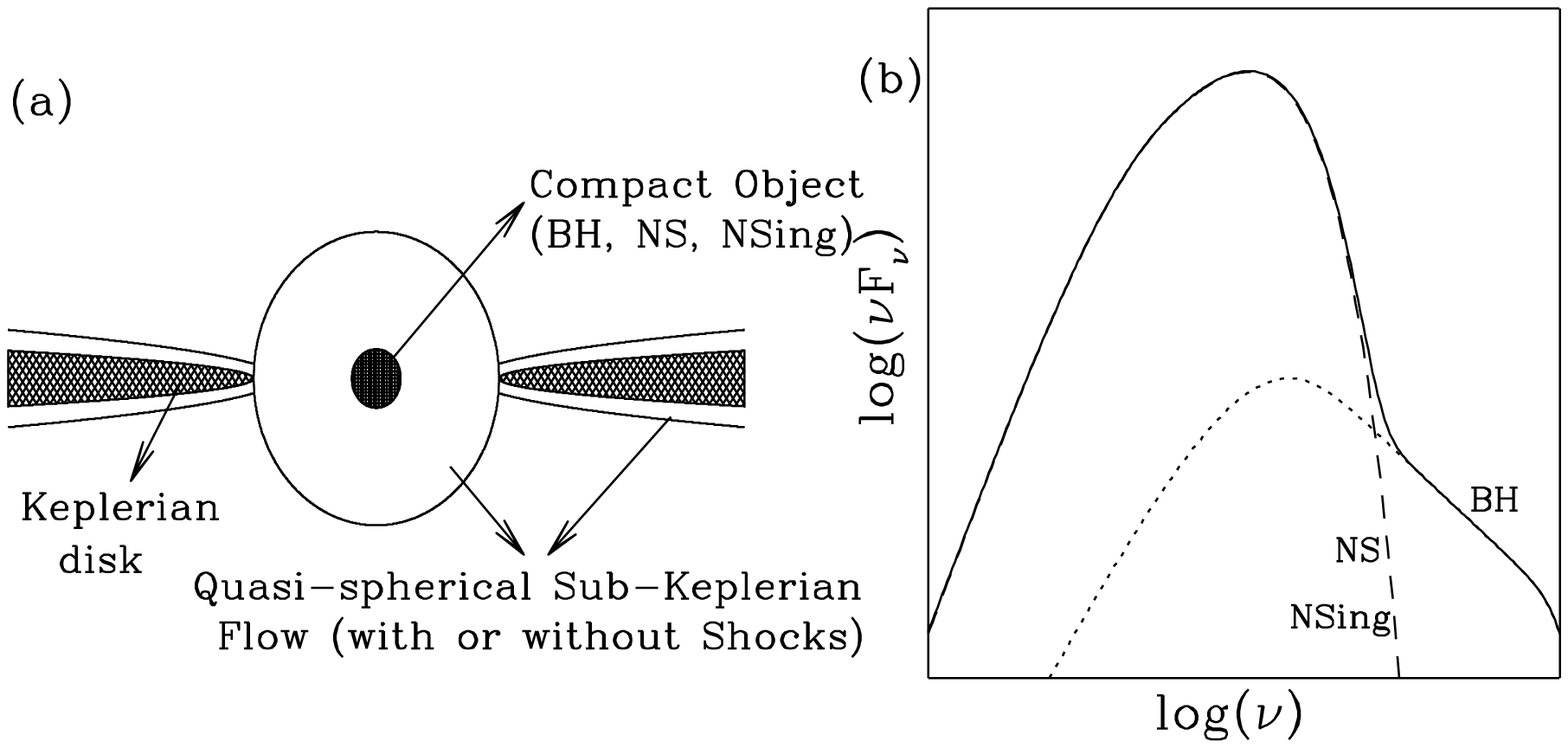,height=13.0truecm,width=13.0truecm,angle=0}}}
\vspace{-8.0cm}
\noindent {\small {\bf Fig. 4:} 
(a) Schematic diagram of the two component advective disk model. Keplerian disk component
which eventually becomes sub-Keplerian close to a compact object is flanked by
a sub-Keplerian halo component which is originated from the Keplerian disk
farther out and possibly  contributed by winds of the companion or nearby stars. 
(b) Soft states of neutron stars (NS), black holes (BH) and naked singularities 
(NSing) are distinguished by the presence or absence of the weak hard tail 
component possibly due to bulk motion Comptonization (Chakrabarti \& Titarchuk, 1995). }
\end{figure}
momentum transfer in thermal Comptonization) and produces
a weak hard tail component with photon index $ \sim 2.5-3$. While on a neutron star
such a hard component must be missing since the flow has to slow down on the inner boundary.
Just for completeness, a naked singularity (NSing, with inner boundary at $x \sim 0$) also
should not have this weak hard tail since extremely dense advecting matter close to the
singularity would carry all such hard photons inwards. This feature is because the
absorbing boundary is at $x=0$ (Chakrabarti \& Titarchuk, 1995). 

{\parfillskip=0pt Fig. 5 shows examples of spectral transitions in black hole candidates 
in all the three models described above. We choose here
$M^*_{BH}=1M_\odot$, which after correction due to spectral hardening (Shimura \& Takahara,
1995), corresponds to a mass of $M_{BH}\sim 3.6 M_\odot$. All the rates are in units of Eddington
rate. In Fig. 5(a), we consider three
disk rates ${\dot m}_d =0.3,\ 0.05,\ 0.0005$ but the same halo rate ${\dot m}_h=1$. Solid, long-dashed
and short-dashed curves are for strong shock Model (TCAFM1), 
weak or no-shock Model (TCAFM2) and zero angular momentum halo  Model (TCAFM3). 
For a set of (${\dot m}_d, {\dot m}_h$), the
spectrum is hardest for TCAFM1 and softest for (TCAFM3). This is expected since
the emission region has the highest density when shocks are stronger. SCAFM
always produces soft states for these parameters.
In Fig. 5(b), we show the comparison of energy spectral 
index $\alpha$ (where $F[E] \sim E^{-\alpha}$) for these models
as functions of ${\dot m}_d$. In all these models, spectra becomes soft
even when the disk rate is much below Eddington rate. In the case of 
supermassive black holes, the behavior is very similar
as the electron temperature of the sub-Keplerian region 
is very insensitive to the central mass ($T_e \propto M_{BH}^{0.04}$). 
At high accretion rate, the bulk motion Comptonization produces 
weaker hard tail (Chakrabarti \& Titarchuk, 1995). Its behavior is independent 
of any model and depends mainly on the optical depth in the last few Schwarzschild 
radii outside the horizon. In the long dashed region of the convergent flow curve,
both power laws due to thermal and bulk motion Comptonizations are expected in the observed
spectra. In Fig. 5(c), the dependence of the spectra on the location where the
flow deviates from a Keplerian disk ($x_K$) is shown using TCAFM3.
Fig. 5(d) shows the corresponding variation of the spectral index.
As described in Chakrabarti \& Titarchuk (1995) and Chakrabarti (1996b), 
this variation of $x_K$ could be simply due to the viscosity variation in the flow
(see, Fig. 2 above) and therefore such \par}
\newpage
\begin{figure}
\vbox{
\vskip -0.6cm
\centerline{
\psfig{figure=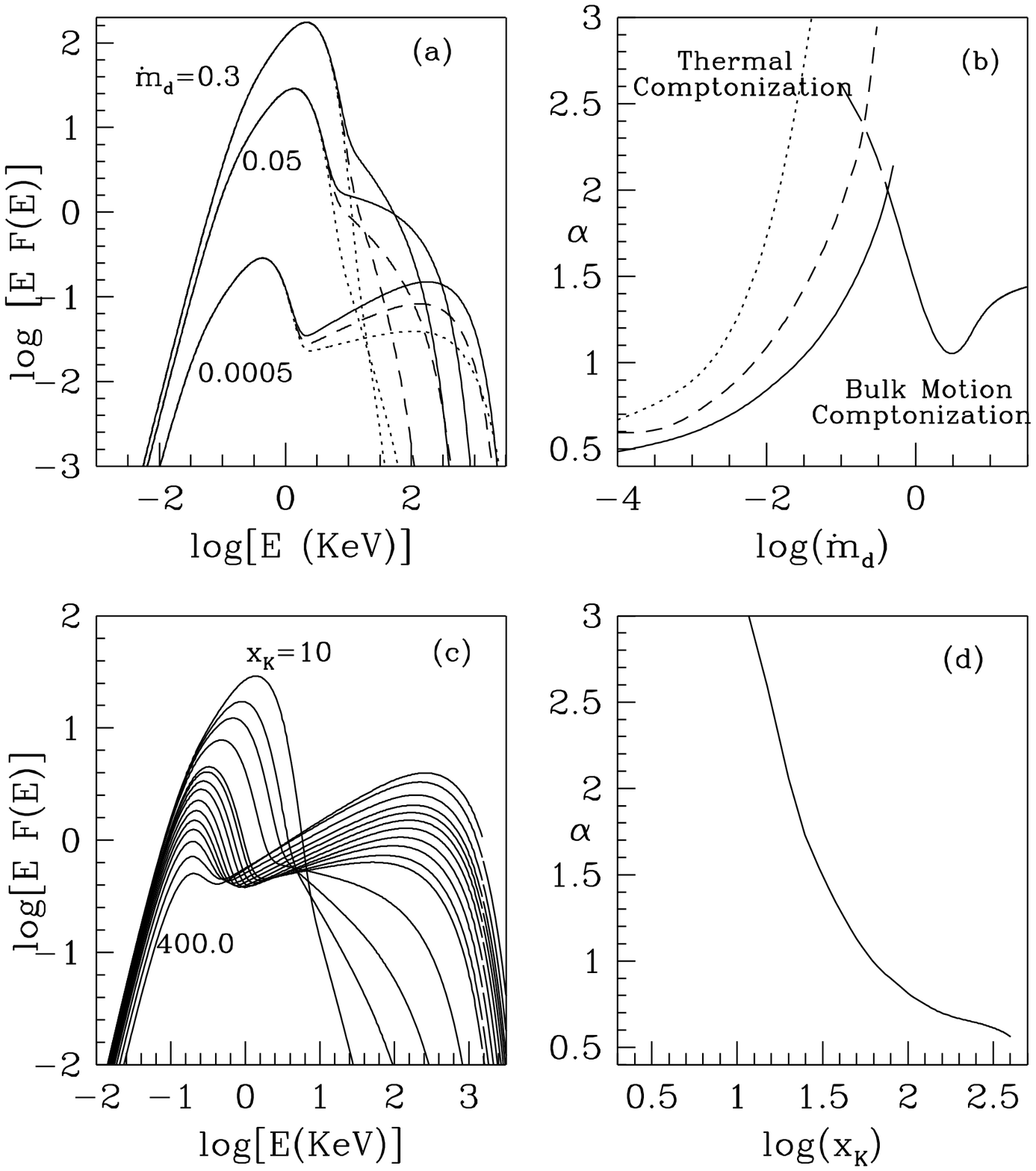,height=13.0truecm,width=13.0truecm,angle=0}}}
\vspace{-2.5cm}
\noindent {\small {\bf Fig. 5:} 
Model dependence of spectral properties. (a) Solid, long-dashed and short-dashed
curves are for TCAFM1, TCAFM2 and TCAFM3 respectively. (b) Spectral indices
of the corresponding models as functions of ${\dot m}_d$ are drawn.
(c) Result of TCAFM3 (${\dot m_d}=0.05,\ {\dot m_h}=1.0$). 
Spectrum changes from hard state to soft state as the Keplerian disk
approaches the black hole due to increase in $\alpha$ and ${\dot m}_d$ (as probably 
in the rising phase of a novae outburst). (d) Changes in spectral index with $x_K$.}
\end{figure}
\noindent variation in the spectra is expected in viscous time scale,
specially during the rising phase of a novae outburst which is presumably induced
by an enhancement in viscosity (see, Cannizzo 1993 and references therein) at
the outer edge of the Keplerian disk. Figs. 5(c-d) were drawn for ${\dot m}_d=0.05$
and ${\dot m}_h=1.0$. Indeed, rising light curves (Fig. 6c-d) 
derived from this consideration is remarkably similar to what is observed.
In reality both rates must change. Such a transition is best understood 
by a numerical simulation of viscous advective flow together with 
radiative transfer (Chakrabarti et al. in preparation). 

It is interesting to compare the results of TCAF Models presented above
with some of the observations of black hole candidates. In Fig. 6 
spectral evolutions of two well known X-ray novae are roughly fitted with TCAFM1. 
The spectral data of GS2000+25 are taken from Tanaka (1991)
and those of GS1124-68 are from Ebisawa (private communication; Ebisawa et al., 1994).
For simplicity and to highlight the similarity between these two objects, 
all the parameters have been kept fixed (with $x_s=10x_g$, $x_g=2GM_{BH}/c^2$,
and Schwarzschild black hole of mass $1M_\odot$ which after correction due to 
spectral hardening corresponds to $3.6M_\odot$) {\it except} for the rates 
${\dot m}_d$ and ${\dot m}_h$.  Figs. 6(a-b) show these fits. From the
derived pair of rates (${\dot m}_d$, ${\dot m}_h$) the intermediate 
rates are interpolated and 
\begin{figure}
\vbox{
\vskip -0.5cm
\centerline{
\psfig{figure=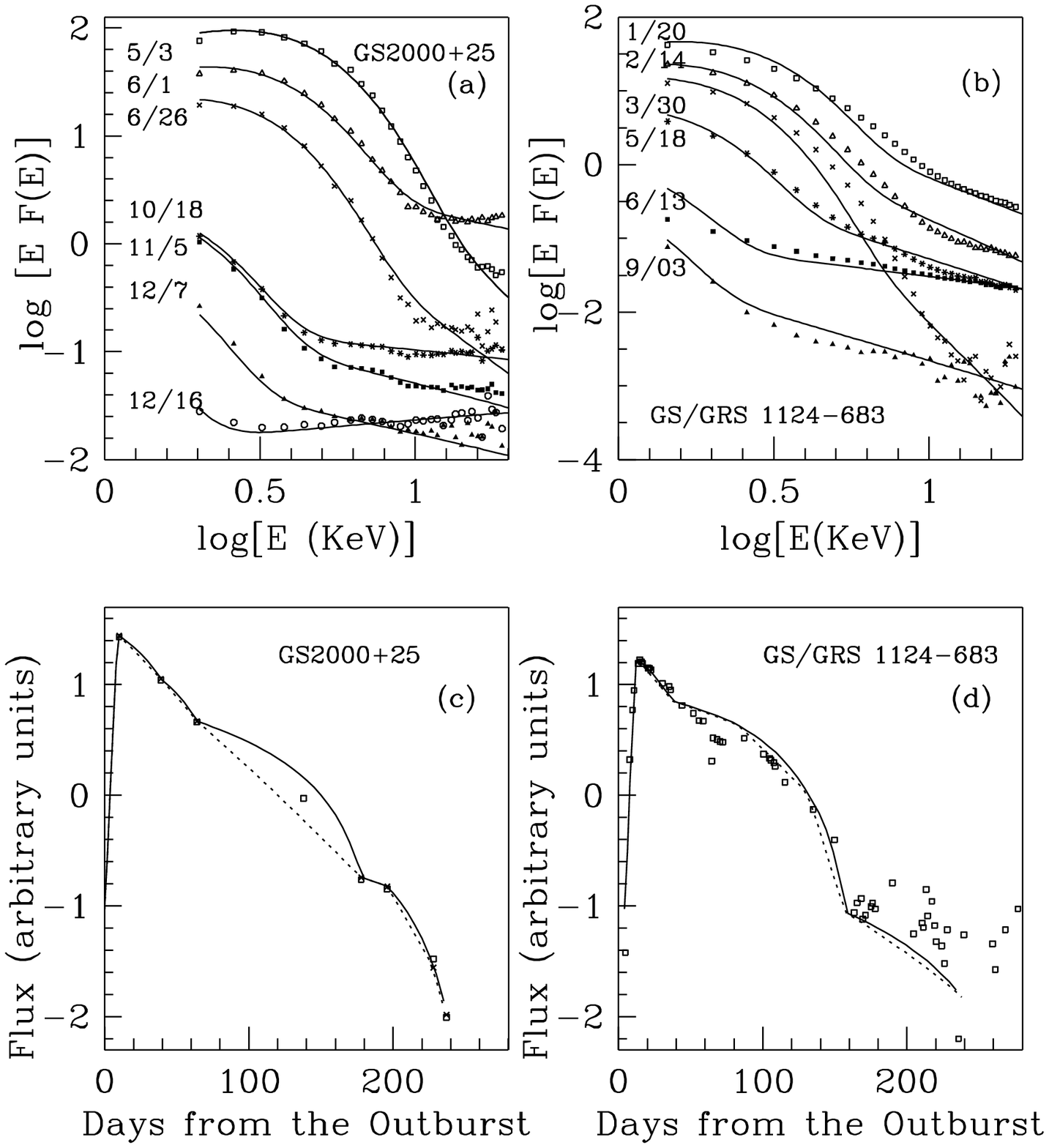,height=13.0truecm,width=13.0truecm,angle=0}}}
\vspace{-2.5cm}
\noindent {\small {\bf Fig. 6:} 
Rough fits of evolution of two X-ray novae spectra using two component advective
flow model (TCAFM1) and comparison of derived light curves with the
observed light curves. See text for details.}
\end{figure}
the resulting light curves 
(1-20keV) are shown in Figs. 6(c-d). Solid and dotted light curves are drawn using
linear-linear and linear-log interpolations of ${\dot m}_d$ respectively. In Fig. 6(c),
squared points are obtained from the actual spectra while crosses are
obtained from the fit in Fig. 6(a). In Fig. 6(d), the squares are from
ASM light curve of Ginga (Kitamoto, private communication; Kitamoto et al.
1992). The general features of the light curves are clearly 
reproduced, including the bumps after 50-70 days 
and another one after 200 days of outburst. Both show a decay time scale
of $\sim 33$days. The bump around 50-70d shows that the decay of disk accretion
rate is not exponential, but more like linear although after around 200days
the disk rate dropped to the point where the exponential decay would have brought it.
The rising light curves in both the cases were computed in two different ways
and the results were similar. In one case (using TCAFM1), the ${\dot m}_d$ was increased
from a quiescent state with an e-folding time of $\sim 2$ days, while in the other
case (using TCAFM3), the $x_K$ was reduced exponentially at a similar rate.

Spectral behavior in above systems and other systems (see, e.g., Ebisawa
et al. 1994 for LMC X-3, Crary et al. 1996 for Cyg X-1), universal
presence of the weak hard tail in soft states, diverse observations
such as quiescent states to rising phases of black hole candidate novae, 
soft to hard transitions, pivoting property of the spectra, quasi-periodic oscillations
(including observed large amplitude modulations) are naturally explained
by TCAF models without invoking any additional unknown components.
Though we did not include magnetic fields explicitly, existence of small
fields as generated by say, Balbus-Hawley instability (this volume),
cannot affect our results. Finally, we wish to point out that even the
jets and outflows are found to be formed more easily from a sub-Keplerian flow
(Chakrabarti \& Bhaskaran, 1992). 

\acknowledgments

The author acknowledges helpful discussions with Drs. Ken Ebisawa, Sunji Kitamoto
and S. Nan Zhang.

\noindent To appear in: `Accretion Phenomena and Related Outflows', proceedings
of 163rd IAU Symposium, July-1996, Eds. D. Wickramsinghe, L. Ferrario
and G. Bicknell.

\noindent Author's address from November 26th, 1996:\\

\noindent Prof. S.K. Chakrabarti\\
\noindent S.N. Bose National Center for Basic Sciences\\
\noindent JD Block, Sector -III, Salt Lake\\
\noindent Calcutta 700091, INDIA\\

\noindent e-mail: chakraba@bose.ernet.in  OR chakraba@tifrc2.tifr.res.in \\ 

\end{document}